\documentclass{iau}
\usepackage{graphicx}

\usepackage{natbib}
\bibpunct{(}{)}{;}{a}{}{,} 
\newcommand{\kms}{{\;\rm km\,s}^{-1}} 
\newcommand{\ksm}{{\;\rm km\;s}^{-1}{\;\rm Mpc}^{-1}}

\renewcommand{\mag}{\mbox{$\;$mag}}
\newcommand{\dex}{\mbox{$\;$dex}}

\title{Allan Sandage and the Distance Scale}

\author[G.A. Tammann \& B. Reindl]{G.A. Tammann \and B. Reindl}

\affiliation{%
     Department of Physics and Astronomy, University of Basel, \\
     Klingelbergstrasse 82, 4056 Basel, Switzerland \\
     email: \texttt{g-a.tammann@unibas.ch}}

\pubyear{2013}
\volume{289}  
\pagerange{1--13}
\jname{Advancing the Physics of Cosmic Distances}
\editors{Richard de~Grijs \& Giuseppe Bono, eds.}
\begin{document}

\maketitle

\begin{abstract}
Allan Sandage returned to the distance scale and the calibration of
the Hubble constant again and again during his active life,
experimenting with different distance indicators. In 1952 his proof of
the high luminosity of Cepheids confirmed Baade's revision of the
distance scale (H$_{0}\sim250\ksm$). During the next 25 years, he
lowered the value to 75 and 55. Upon the arrival of the
\textit{Hubble Space Telescope}, he observed Cepheids to calibrate
the mean luminosity of nearby Type Ia supernovae (SNe\,Ia) which, 
used as standard candles, led to the cosmic value of 
H$_{0}=62.3\pm1.3\pm5.0\ksm$. Eventually he turned to the tip of the
red-giant branch (TRGB) as a very powerful distance indicator. A
compilation of 176 TRGB distances yielded a mean, very local value of
H$_{0}=62.9\pm1.6\ksm$ and shed light on the streaming velocities in
the Local Supercluster. Moreover, TRGB distances are now available for
six SNe\,Ia; if their mean luminosity is applied to distant SNe\,Ia,
one obtains H$_{0}=64.6\pm1.6\pm2.0\ksm$. The weighted mean of the two
\textit{independent} large-scale calibrations yields H$_{0}=64.1\ksm$
within 3.6\%.   

\end{abstract}

\firstsection 

\section{Introduction}
\label{sec:1}
Allan Sandage (1926-2010) was one of the best observers ever, but
his driver was always the physical understanding of astronomical
phenomena. Physics as his primary goal is exemplified in his
ground-braking paper \textit{`The ability of the 200-inch Telescope to
discriminate between selected World Models'} \citeyearpar{Sandage:61}
that became the basis of modern observational cosmology. 
Other examples are the emerging understanding of stellar evolution
\citep{Sandage:Schwarzschild:52}, the theory of stellar pulsation
\citep{Sandage:58a,SBT:99}, the formation of the Galaxy \citep{ELS:62}
and of other galaxies \citep{Sandage:86}, including their violent
stages \citep{BBS:63}, the nature of the expansion field
\citep{HMS:56,Sandage:75,SRT:10}, and the age dating of stellar
clusters \citep[e.g.,][]{Sandage:58b} as well as of the Universe
\citep[e.g.,][]{Sandage:70,Sandage:72}. He devoted several papers to
the Tolman test of the nature of redshifts \citep[e.g.,][]{Sandage:10}
and he is the father of what has become known as the Sandage-Loeb
effect \citep{Sandage:62a}.  

     Walter Baade, the thesis adviser of Sandage, gave him a tough
training in the ancient and intricate art of observing; it was later
described in Sandage's outstanding history,
\textit{`The Mount Wilson Observatory'} \citeyearpar{Sandage:04}. 
Soon Sandage excelled himself in that high
art and was commissioned to observe also for Hubble. He loved to
observe on Mount Wilson and Palomar Mountain. After the split of the
Carnegie Institution and the California Institute of Technology in
1980, he did not go to Palomar again. On Mount Wilson he was one of
the last observers before the Mountain was given into other hands in
the mid-1980s. A consolation became the wide-field $2.5\;$m telescope
of the Las Campanas Observatory, where he took many direct plates for
his atlases and for the extensive Virgo Cluster survey \citep{SBT:85}.
In total, he has spent more than 2000 nights at various telescopes.
He felt a strong responsibility to reduce and publish the enormous
amount of accumulated observations. After 1990 he used only the 
\textsl{Hubble Space Telescope (HST)}; it was Abhijit Saha who
introduced him to the novel technique of CCD photometry.

\section{Early Work}
\label{sec:2}
\begin{figure}
\begin{center}
\begin{minipage}[t]{.58\linewidth}
   \vspace{0pt}
   \hspace*{-0.2cm}
   \includegraphics[width=1.00\textwidth]{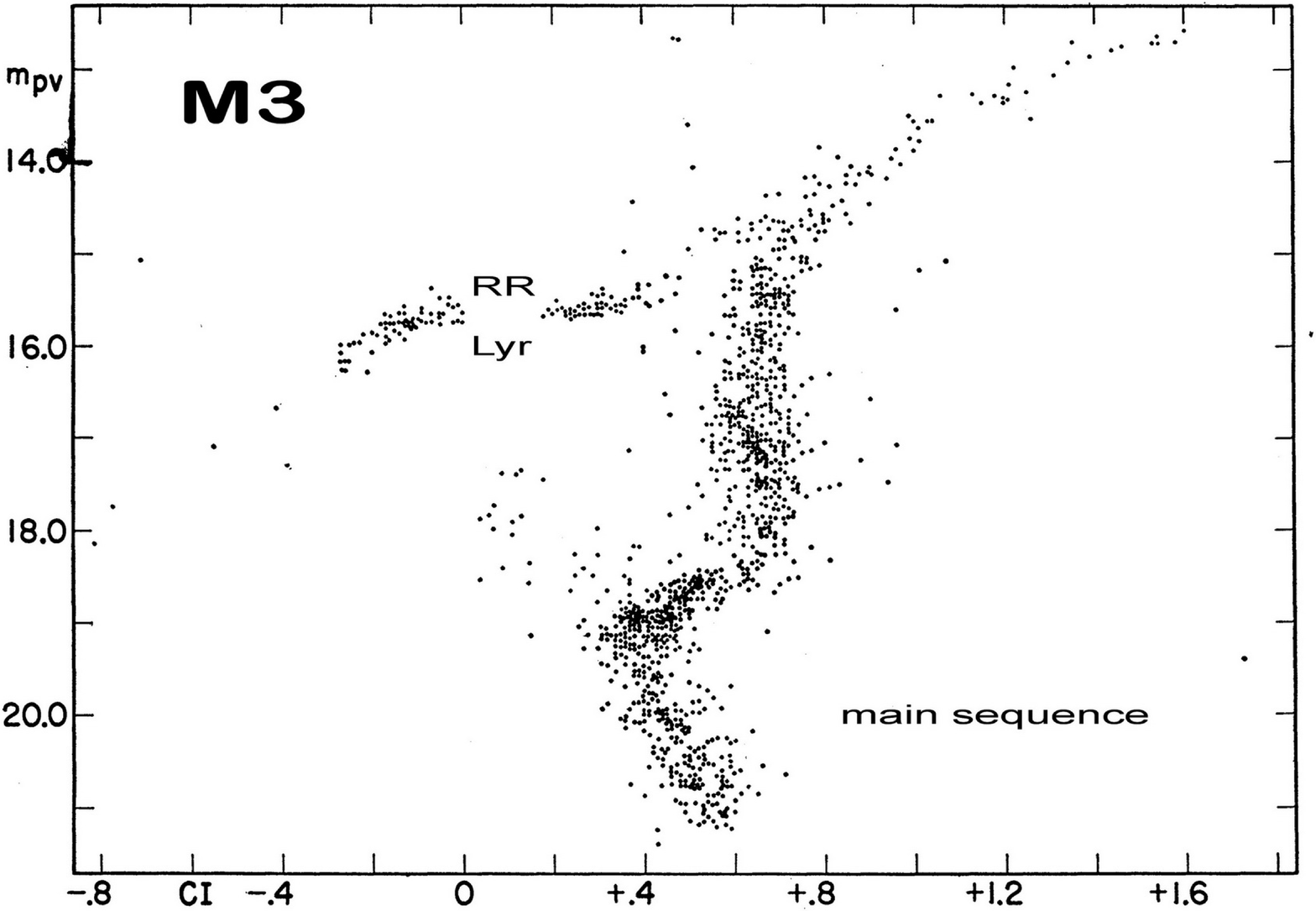}
\end{minipage}
\begin{minipage}[t]{.40\linewidth}
   \vspace{0pt}
   \hspace*{0.00cm}
   \includegraphics[width=1.00\textwidth]{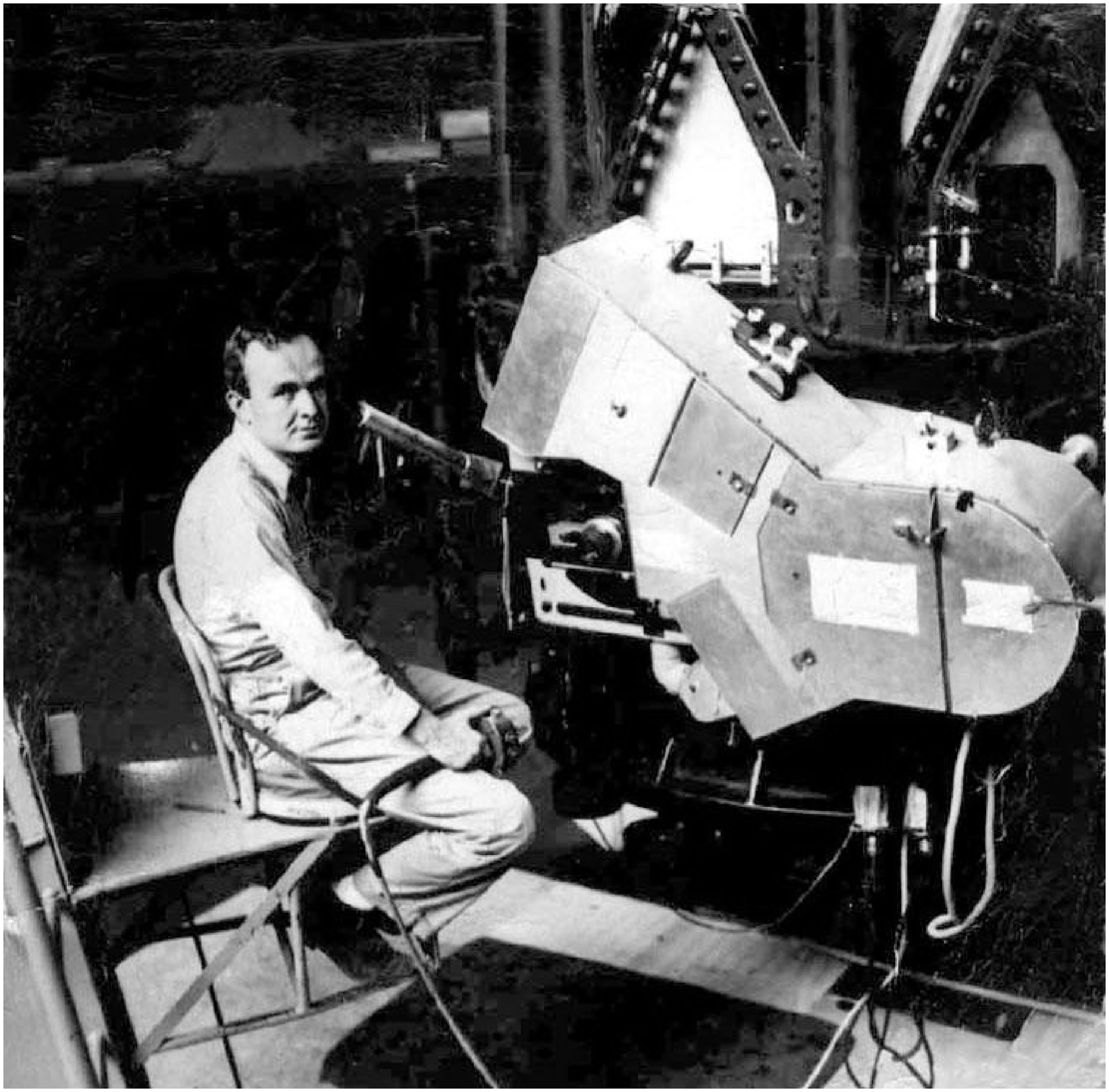}
\end{minipage}
\end{center}
   \caption{%
     (left)
     Color-magnitude diagram (CMD) of the globular cluster M3 from
     Sandage's thesis, where he matched the cluster main sequence with
     the main sequence of the young Population~I, showed the locus of
     RR Lyrae stars, and discussed the influence of metallicity on the
     CMD.
     (right)
     Allan Sandage at the spectrograph of the Mount Wilson 60-inch
     telescope in 1950.  
}   
   \label{fig:01}
\end{figure}

His thesis assignment was the color-magnitude diagram (CMD) of the
globular cluster M3. By pushing the photometry down to an
unprecedented limit of $23^{\rm rd}\mag$, he could find the cluster's
main sequence and fit it to the main sequence of the young
Population~I \citep[Fig.\,\ref{fig:01};][]{Sandage:53}.  
This provided an essential clue to the theory of stellar evolution,
but at the same time it solved a long-standing problem of the distance
scale. \citet{Baade:44} had pointed out that galaxy distances derived
from RR~Lyrae stars and Cepheids were in contradiction. It became now
clear that the RR~Lyrae calibration was roughly correct, whereas the
Cepheids were too faint by $\sim\!1.5\mag$. This confirmed
\citet*{Baade:44} proposal that all of Hubble's distances had to be
doubled. In the following more than 50~years Sandage published some
40 papers on the physics and luminosity of RR~Lyrae stars and 50
papers on Cepheids. The distance scale runs like a red line through
his entire professional life. 

     The next stretch came with the paper of \citet*{HMS:56}, for
which Sandage had written the analytical part. He much improved
Hubble's galaxy magnitudes and showed that what Hubble had identified
as brightest stars in other galaxies were in fact \textsc{Hii}
regions. This led to H$_{0}=180\pm40\ksm$. Two years later, he
\citep{Sandage:58a} concluded from a re-discussion of the Cepheid
distances and from straightening out the confusion of \textsc{Hii}
regions and brightest stars that $50<{\rm H}_{0}<100\ksm$; this was an
increase of Hubble's distance scale by a factor 5 to 10.
\citet{Sandage:62b} defended this range of H$_{0}$, also using the
angular size of the largest \textsc{Hii} regions and novae, against
higher values proposed at the influential Santa Barbara Conference on
Extragalactic Research.  

     Determination of the extragalactic distance scale had been
described by \citet{Baade:48} as one of the main goals of the
forthcoming 200~inch Telescope. Correspondingly, Baade and Hubble had
taken repeated photographs, once the telescope had gone into
operation, of several galaxies for work on their Cepheids; they were
joined by Sandage in the following years. A parallel task was for
Sandage, W.~A. Baum, and H.~C. Arp to establish reliable photoelectric
magnitude sequences in a number of fields. In the mid-1960s, there
were sufficient data for Sandage, who had inherited the photographic
plates of Baade and Hubble, to start a new onslaught on H$_{0}$. 

\begin{figure}
\begin{center}
\begin{minipage}[t]{.63\linewidth}
   \vspace{0pt}
   \hspace*{-0.4cm}
   \includegraphics[width=1.00\textwidth]{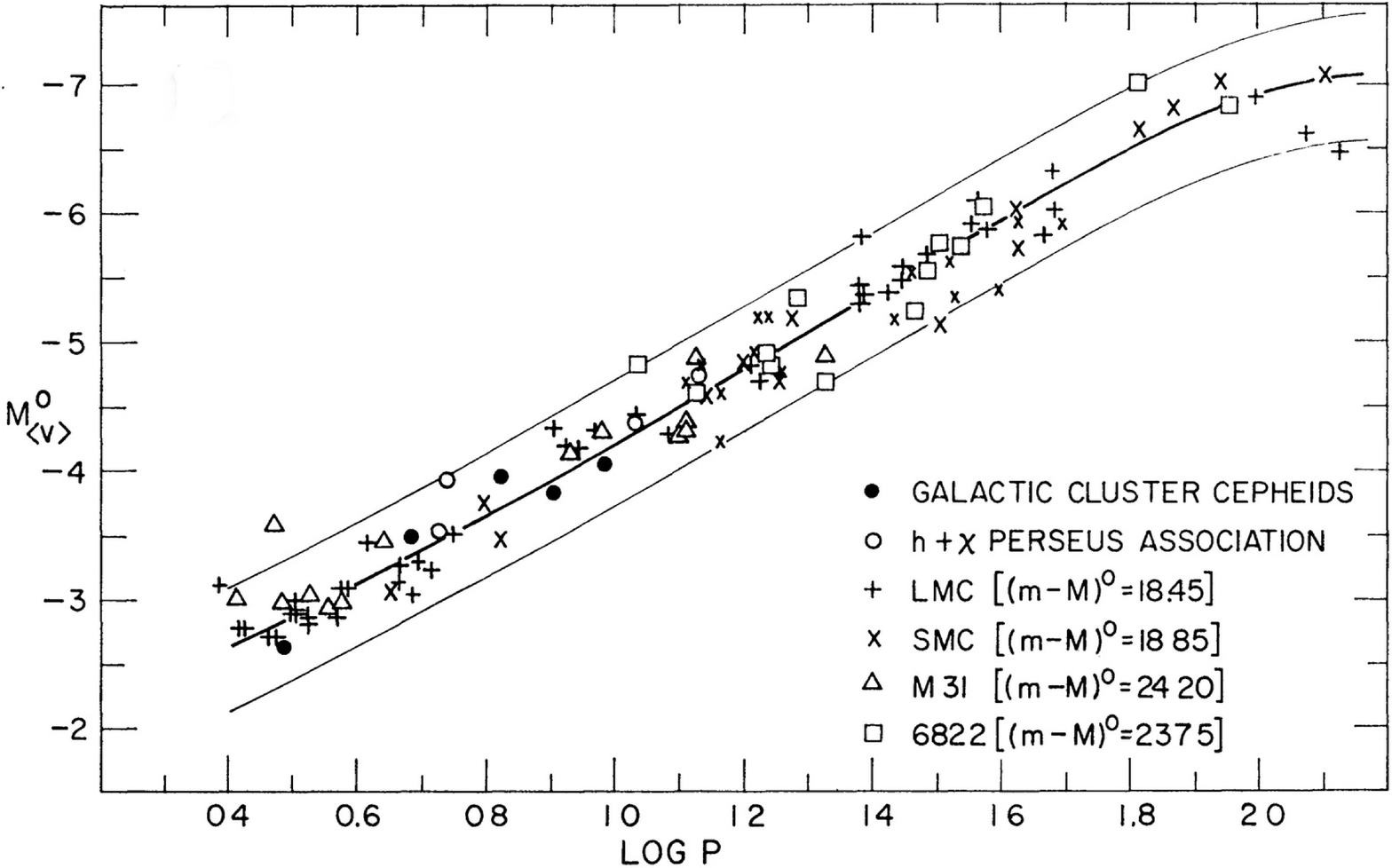}
\end{minipage}
\begin{minipage}[t]{.32\linewidth}
   \vspace{0pt}
   \hspace*{0.10cm}
   \includegraphics[width=1.00\textwidth]{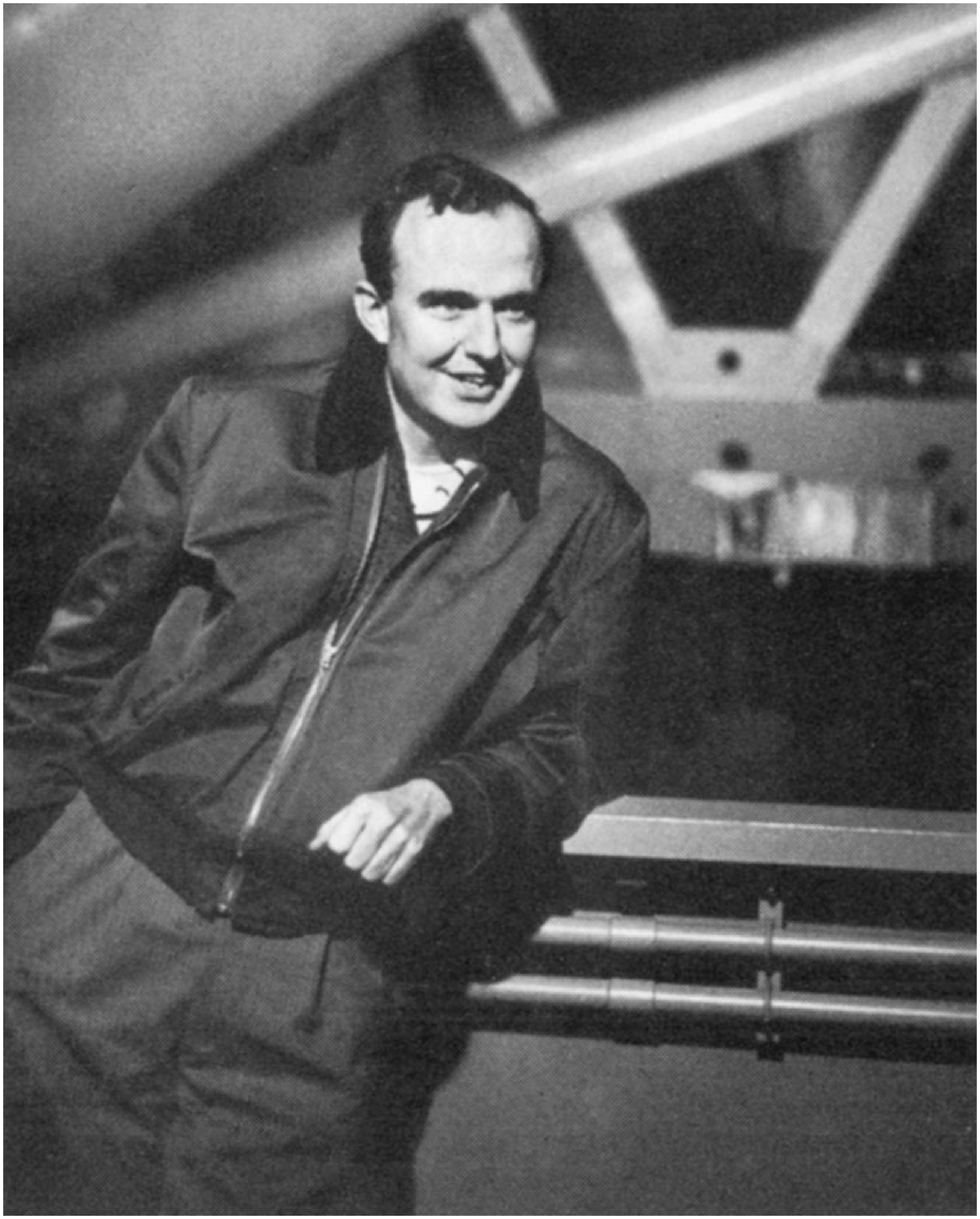}
\end{minipage}
\end{center}
   \caption{%
     (left)
     The P-L relation in 1968 from a superposition of Cepheids in four
     galaxies and five Galactic Cepheids with known distances. The
     resulting distances are marked.
     (right)
     Allan Sandage in front of the 200~inch Telescope on Palomar
     Mountain in 1953.
}   
   \label{fig:02}
\end{figure}

     It was preceded by a new period-luminosity (P-L) relation for
Cepheids (Sandage \& Tammann \citeyear{ST:68}), which was a
superposition of the known Cepheids in Large and Small Magellanic
Clouds (LMC, SMC), M31, and NGC\,6822, and whose zero point was based
on five Galactic Cepheids that are members of open clusters with known
distances (see Fig.\,\ref{fig:02}). The moduli of the LMC and SMC
agree with modern values to  within $0.1\mag$, the modulus of M31,
based on the excellent Cepheids of \citet{Baade:Swope:63}, was still
$0.2\mag$ short, and the distance to NGC\,6822 was too long due to
internal absorption. Another prerequisite for the new determination of
H$_{0}$ was the first Cepheid distance outside the Local Group, i.e.\
of the highly resolved galaxy NGC\,2403 \citep{TS:68}, which was taken
as representative for the whole M81 group.

\section{The series `Steps toward the Hubble Constant'}
\label{sec:3}
\begin{figure}
\begin{center}
\begin{minipage}[t]{.58\linewidth}
   \vspace{0pt}
   \hspace*{-0.5cm}
   \includegraphics[width=1.00\textwidth]{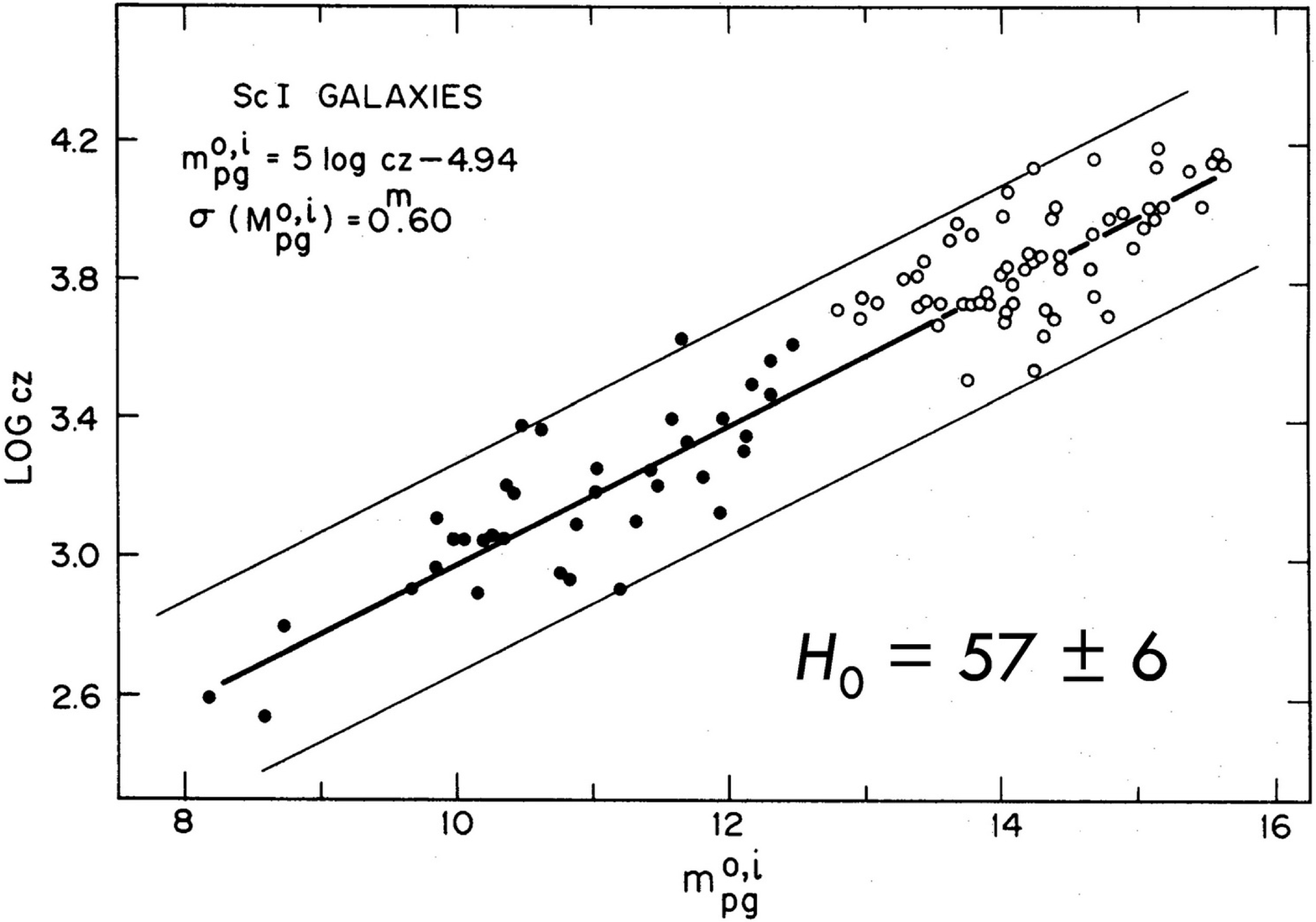}
\end{minipage}
\begin{minipage}[t]{.32\linewidth}
   \vspace{0pt}
   \hspace*{0.40cm}
   \includegraphics[width=1.00\textwidth]{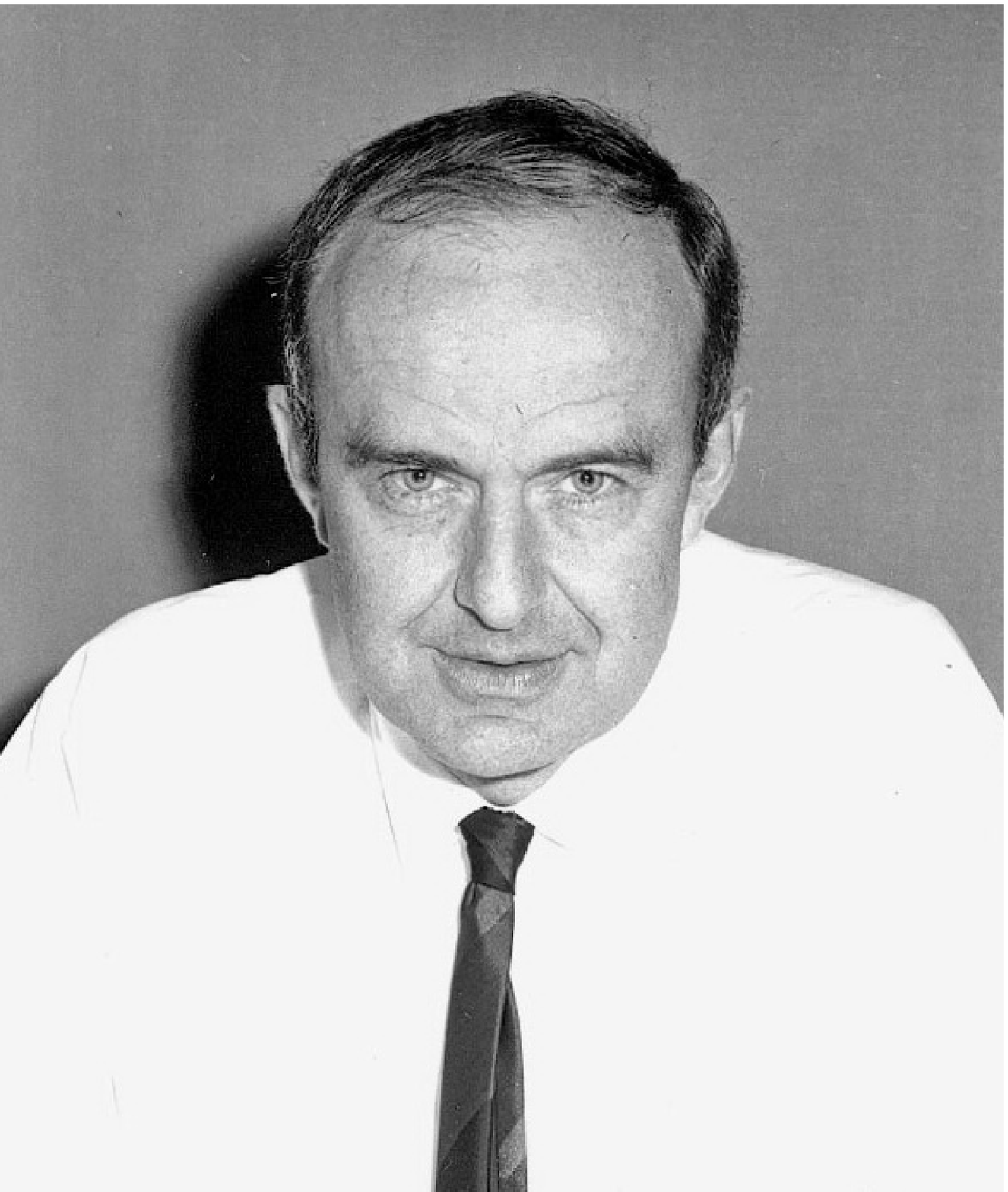}
\end{minipage}
\end{center}
   \caption{%
     (left)
     Hubble diagram of luminosity class\,I spirals. Filled symbols are
     Sc\,I galaxies from the Shapley-Ames Catalog, open symbols are
     newly selected SC\,I galaxies. Plotted is $\log v$ versus the
     corrected Zwicky magnitude.
     (right)
     Allan Sandage (1967).
}   
   \label{fig:03}
\end{figure}

In the two first steps of the series \citep{ST:Steps}, the linear size
of the largest \textsc{Hii} regions and the luminosity of the
brightest blue and red stars were calibrated with the available
Cepheid distances. The Maximum \textsc{Hii}-region size and blue-star
luminosity were found to increase with galaxy size, whereas the
red-star luminosity is quite stable. In step~III, the results were
applied to the M101 group which was found at a then surprisingly large
distance of $(m-M) \ge 29.3$, yet in agreement with the modern value
(see Section~\ref{sec:7}). Step~IV served to extend and calibrate
\citet{vandenBergh:60} luminosity classes of spiral and irregular
galaxies. These morphological classes depend on the surface brightness
and the `beauty' of the spiral structure. In particular, it was found
that giant Sc spirals (Sc\,\textsc{i}) have a mean intrinsic face-on
luminosity of $M_{\rm pg}=-21.25\pm0.07\mag$. The additional inclusion
of fainter luminosity classes gave a Virgo Cluster modulus of
$(m-M)=31.45\pm0.09\mag$, which, with an adopted cluster velocity of 
$1111\kms$, led to a first hint of H$_{0}=57\pm6\ksm$.  
To extend the distance scale farther out, remote Sc\,\textsc{i}
spirals were selected in the Polar Caps of the National
Geographic-Palomar Sky Survey in step~VI. 
For 69 of these galaxies, spectra could be measured, resulting in
velocities of $2700<v<21,000\kms$. Their apparent magnitudes,
$m_{\rm pg}$, were taken from the Zwicky Catalog \citep{Zwicky:Catalog}
and corrected for Galactic and internal absorption.
The sample out to $15,000\kms$ is shown in the Hubble diagram in
Fig.~\ref{fig:03}; it is necessarily biased by the magnitude limit of
the catalog. A subsample of 36 galaxies within $8,500\kms$ was
therefore isolated. It was ensured that is was not affected by the
magnitude limit. This sample and the above luminosity calibration
yields H$_{0}=56.9\ksm$ with a statistical error of $\pm3.4\ksm$.  

     The summary paper (numbered as step~V) gives H$_{0}=57\ksm$ with
a small statistical error for the global value and with a tacit
understanding that the systematic error is on the order of 10\%. The
\textit{mean} expansion rate of the nearby galaxies was shown not to
depend on the Supergalactic longitude and also to be independent of
distance, in agreement with an earlier conclusion of \citet{STH:72}.
The average random velocity of a nearby galaxy was found to be
$\sim\!50\kms$. 

\section{Selection Effects}
\label{sec:4}
The results of the `Steps Toward the Hubble Constant' met with stiff
opposition from G.~de Vaucouleurs for the following 10 years. He
considered the expansion rate to be a stochastic variable, depending
on distance and direction with a mean value of $90<H_{0}<110\ksm$
\citep[e.g.,][]{deVaucouleurs:Peters:85}. The main reason for the
disagreement is not that his very local galaxies were $\sim\!0.3\mag$
closer than adopted by Sandage and than modern values, but that he did
not distinguish between magnitude-limited and distance-limited samples. 

\begin{figure}
\begin{center}
\begin{minipage}[t]{.69\linewidth}
   \vspace{0pt}
   \hspace*{-0.25cm}
   \includegraphics[width=1.00\textwidth]{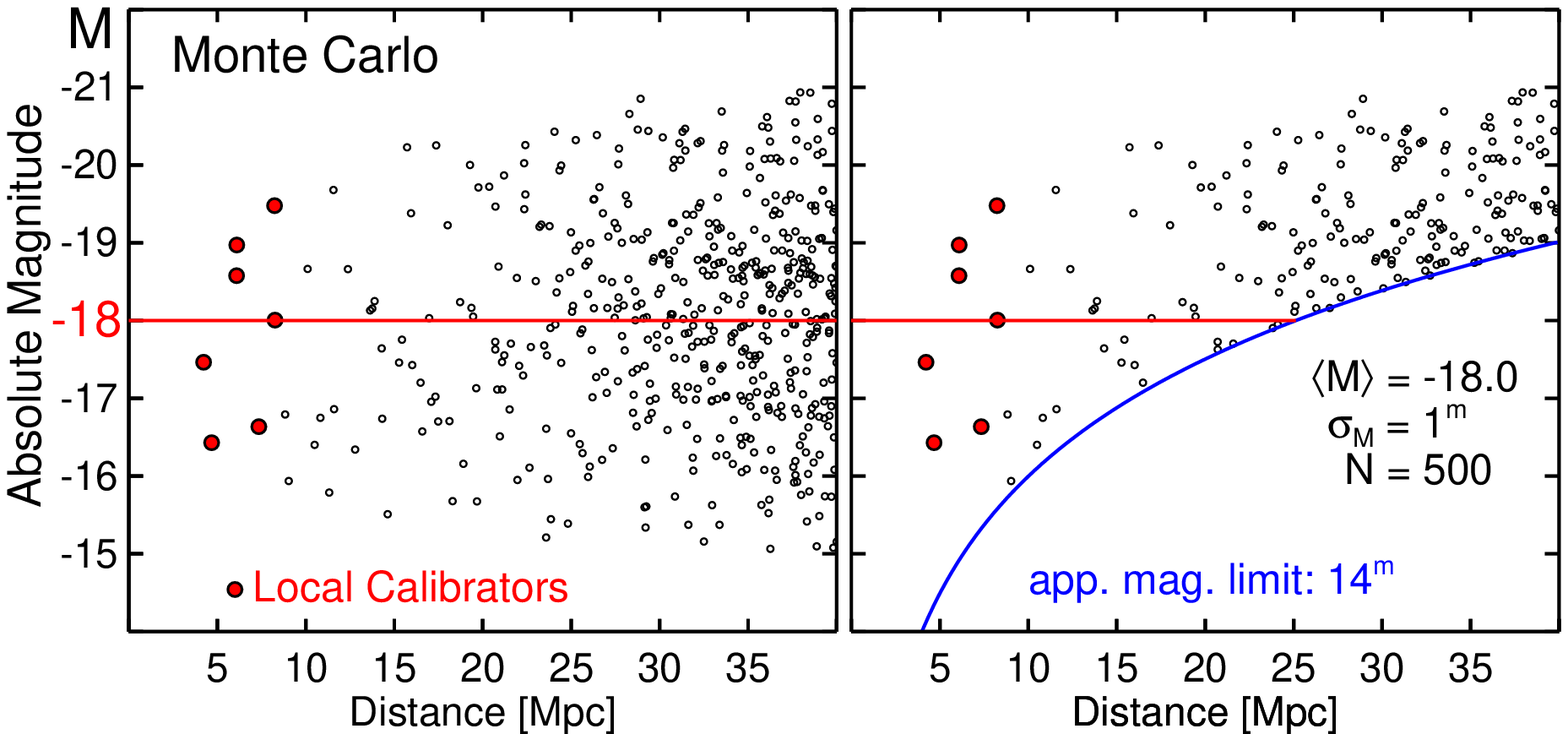}
\end{minipage}
\begin{minipage}[t]{.29\linewidth}
   \vspace{2.5pt}
   \hspace*{0pt}
   \includegraphics[width=1.00\textwidth]{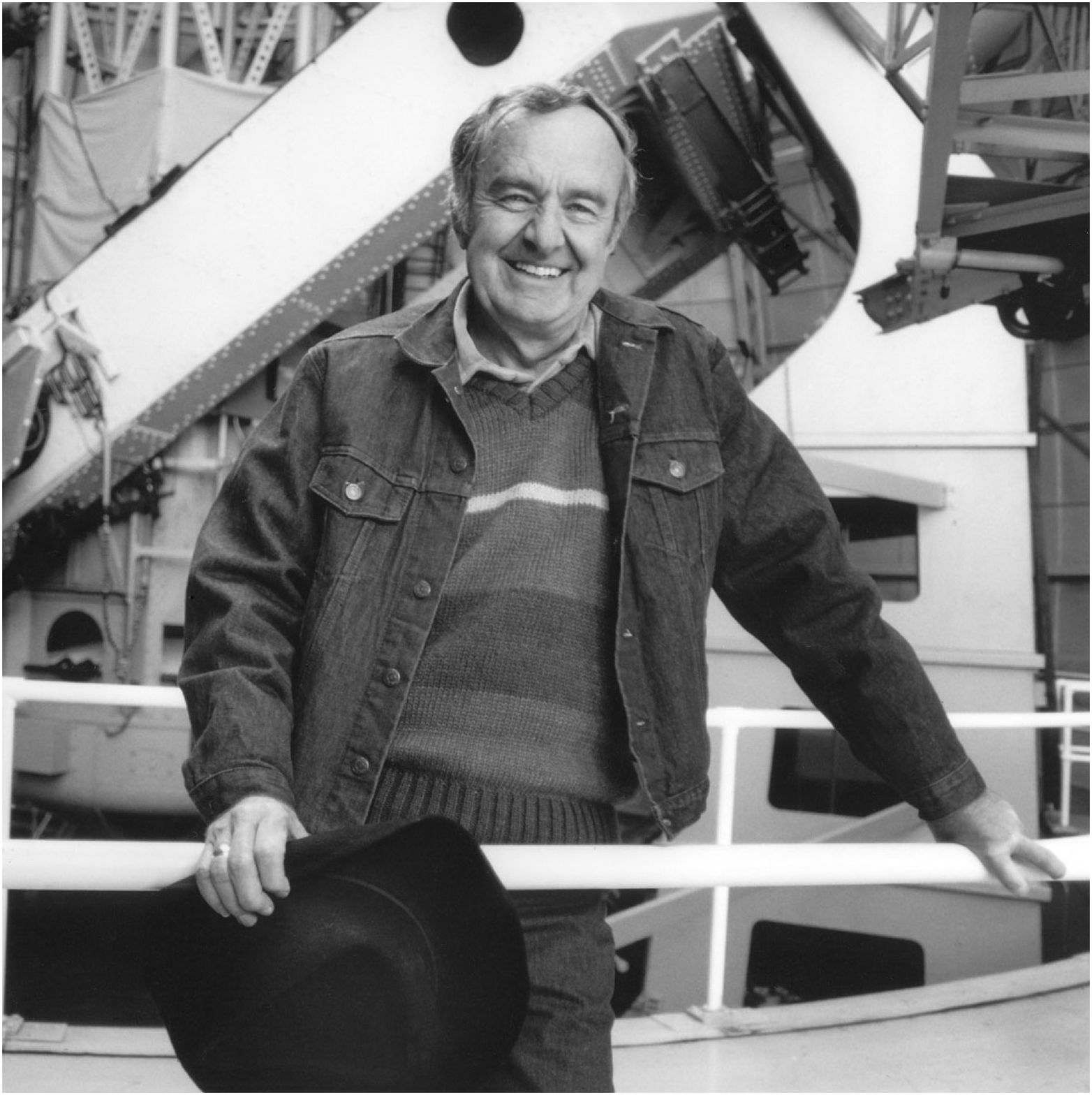}
\end{minipage}
\end{center}
   \caption{%
     (left)
     Monte Carlo distribution of a galaxy sample with the properties
     described in the text. 
     (middle)
     The same distribution, but cut by an apparent-magnitude limit.
     The mean luminosity of the remaining sample increases with
     distance. A few local calibrators are shown as open circles.
     (right)
     Allan Sandage in front of the 100~inch telescope on Mount Wilson.
     \textit{(Photo Douglas Carr Cunningham, 1982)}  
}   
   \label{fig:04}
\end{figure}

     The distinction is decisive in cases where the distance indicator
has non-negligible intrinsic scatter, as illustrated in
Fig.~\ref{fig:04}. It shows a Monte Carlo distribution of 500 galaxies
randomly distributed within 40\,Mpc; their absolute magnitude is
$-18\mag$, with an intrinsic dispersion of $1\mag$. As long as the
sample is complete to the given distance limit, their mean luminosity
does not change with distance. If the same galaxies are cataloged with
a limiting magnitude of $m=14\mag$, then the less luminous galaxies are
progressively excluded; the rest of the sample has unfortunate
statistical properties: the mean luminosity increases with distance,
in the present sample by as much as $1\mag$ and the apparent dispersion
decreases with distance. The small dispersion at large distances has
often led to the erroneous conclusion that the true scatter is small
and that selection bias was negligible.  

\begin{figure}
\begin{center}
\begin{minipage}[t]{.50\linewidth}
   \vspace{0pt}
   \hspace*{-0.4cm}
   \includegraphics[width=1.00\textwidth]{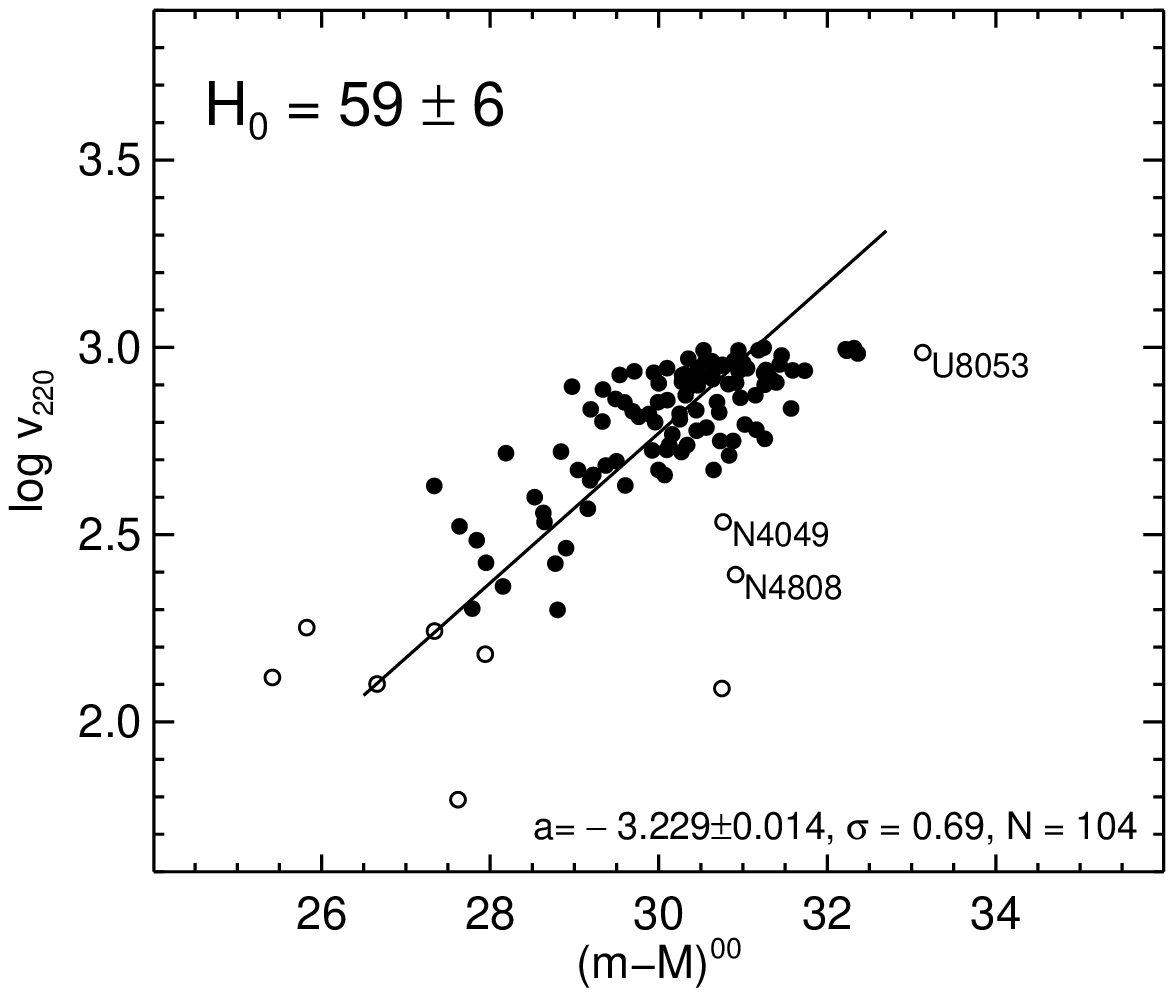}
\end{minipage}
\begin{minipage}[t]{.40\linewidth}
   \vspace{8pt}
   \hspace*{7pt}
   \centering
   \includegraphics[width=1.00\textwidth]{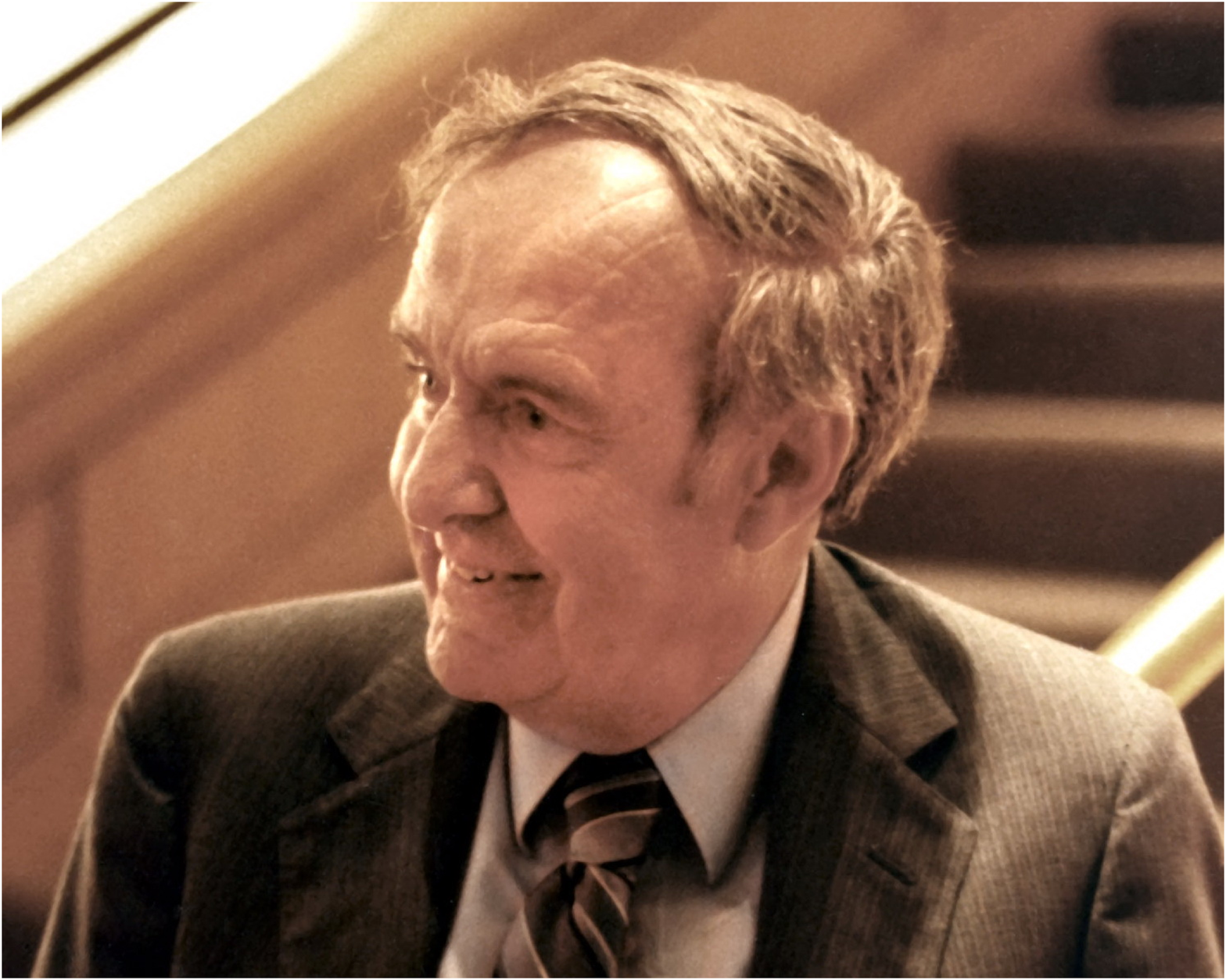}
\end{minipage}
\end{center}
   \caption{%
     (left) 
     Hubble diagram with the T-F distances of an almost complete
     sample of 104 inclined spiral galaxies within  $1000\kms$. 
     (right)
     Allan Sandage in Baltimore (1988). 
}   
   \label{fig:05}
\end{figure}

 Selection bias is particularly dangerous for the Tully-Fisher (T-F) 
relation because of its large intrinsic scatter of $\sim\!\!0.5\mag$.
Applications of the method to magnitude-limited samples that are not
even complete has notoriously led to overestimated values of H$_{0}$.
A Hubble diagram of a \textit{distance-limited}, almost complete
sample of T-F distances, necessarily limited to $v_{220}<1000\kms$,
gives a local value of H$_{0}=59\pm6\ksm$
(see Fig.\,\ref{fig:05}; \citealt{TSR:08b}).  

     Sandage has written a number of papers on various forms of
selection bias and its compensation in favorable cases, including a
series of 12 papers \citep[e.g.,][]{STF:95,Sandage:00}.

\section{The \textsl{HST} Project for the Luminosity Calibration of SNe\,Ia}
\label{sec:5}
\begin{figure}
\begin{center}
\begin{minipage}[t]{.70\linewidth}
   \vspace{18pt}
   \hspace*{-0.2cm}
   \includegraphics[width=1.00\textwidth]{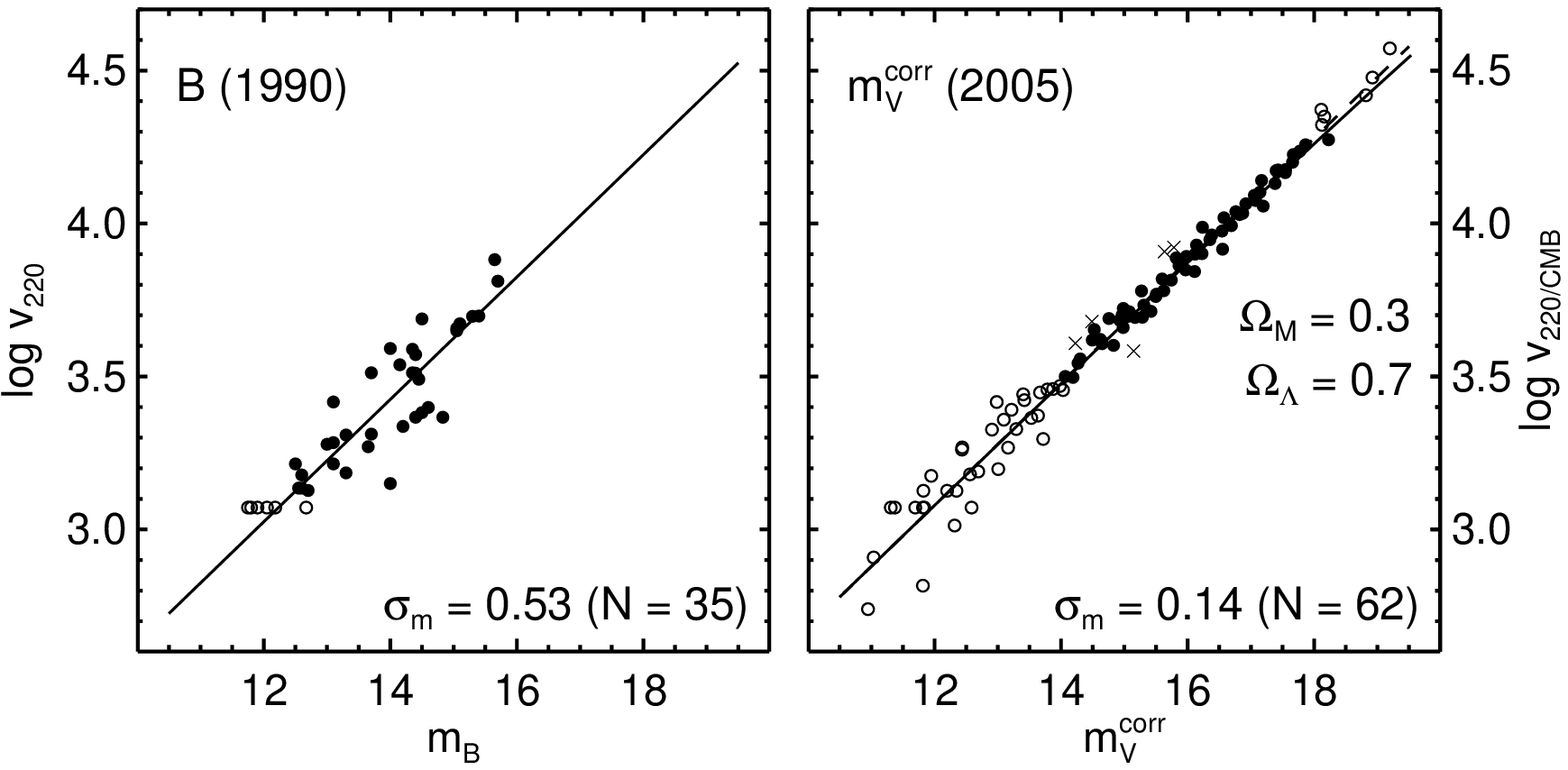}
\end{minipage}
\begin{minipage}[t]{.29\linewidth}
   \vspace{10pt}
   \centering
   \includegraphics[width=1.00\textwidth]{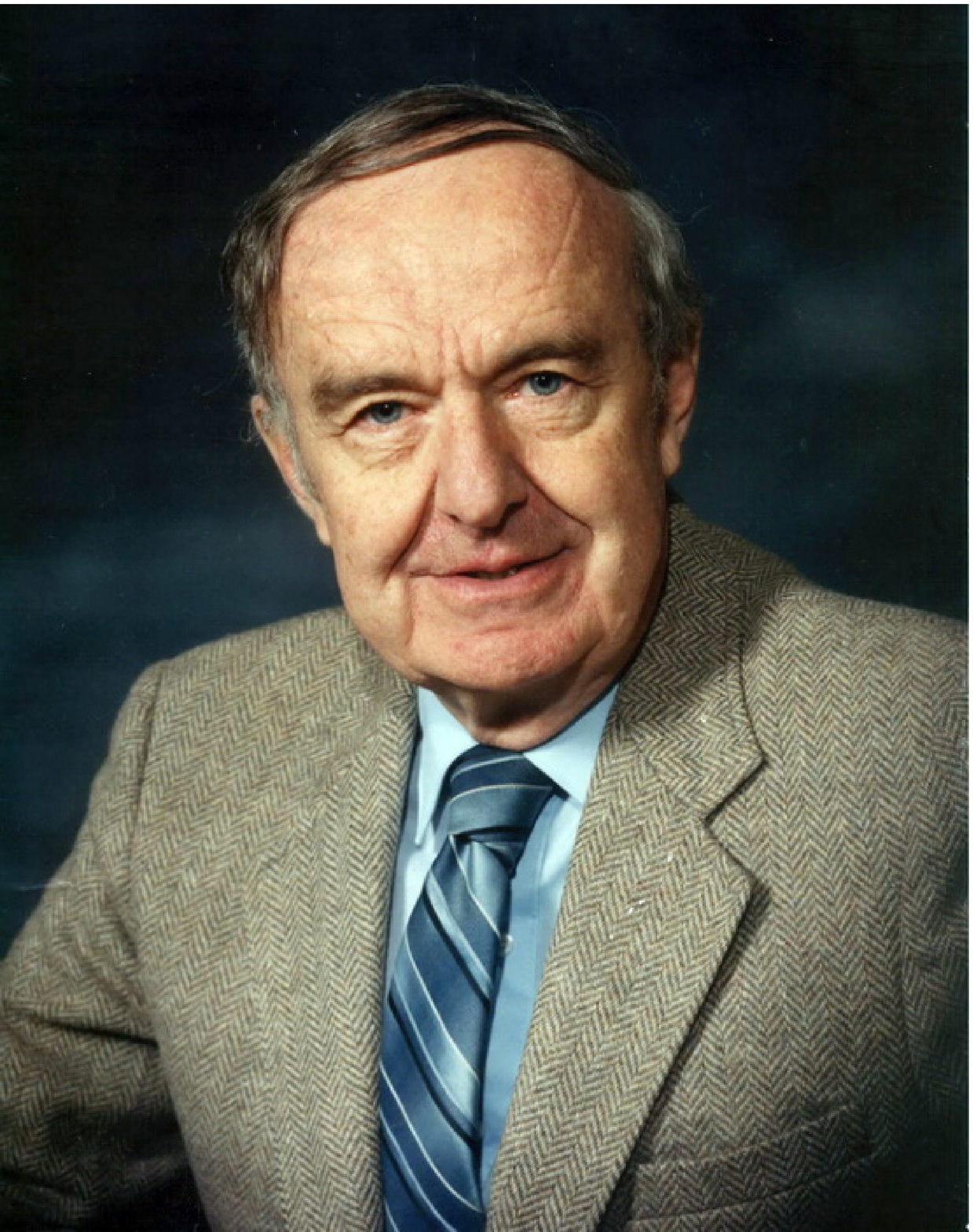}
\end{minipage}
\end{center}
   \caption{%
     (left)
     Hubble diagram of SNe\,Ia as of 1990 \citep{TL:90}; it was
     limited to $\sim\!10,000\kms$ and had large scatter.
     (middle)
     Same, as of 2005 \citep{RTS:05} out to $30,000\kms$. The 62
     SNe\,Ia with $3000<v<20,000\kms$ (black dots) have a scatter of
     only $0.14\mag$. The slightly curved Hubble line assumes
     $\Omega_{M}=0.3$ and $\Omega_{\Lambda}=0.7$.   
     (right)
     Allan Sandage in Baltimore (1988). 
}   
   \label{fig:06}
\end{figure}

During the planning phase for \textsl{HST}, Sandage formed a small
team, including A.~Saha, F.~D.~Macchetto, N.~Panagia, and
G.~A.~Tammann, to determine the Cepheid distances of some galaxies
with known Type Ia supernovae (SNe\,Ia), which---with their maximum
magnitudes taken as standard candles---should lead to a large-scale
value of H$_{0}$. In spite of a pilot paper \citep{ST:82}, the small
luminosity dispersion of SNe\,Ia was not yet established in 1990 
(see Fig.~\ref{fig:06}, left), but they were proven as exquisite
standard candles in the following years.  
The SN \textsl{HST} Project was complimentary to the \textsl{HST} Key
Project for H$_{0}$ that originally did not include SNe\,Ia as
targets. 

     The maximum magnitude of SNe\,Ia must be corrected for internal
absorption and standardized to a fixed decline rate $\Delta m_{15}$
\citep{Phillips:93}. Different procedures have been proposed. 
We use here the $m_{V}^{\rm corr}$ magnitudes of \citet{RTS:05}
because they define a Hubble diagram with a particularly small
scatter of $0.14\mag$ (see Fig.~\ref{fig:06} middle) and yield a 
well-defined intercept of $C_{V}=0.688\pm0.004$. It should be
emphasized that the corrected magnitudes of other authors may not have
the same zero point because of different choices of intrinsic colors,
absorption laws, and reference values of $\Delta m_{15}$.  

     The intercept $C_{V}$ of the Hubble line is defined as 
$C\equiv\log {\rm H}_{0}-0.2M-5$, hence in the present case,
\begin{equation}
   {\rm H}_{0}=0.2 \langle M_{V}^{\rm corr}\rangle + (5.688\pm0.004),
\label{eq:01}
\end{equation}
where $\langle M_{V}^{\rm corr}\rangle$ is the mean absolute magnitude
of SNe\,Ia. The resulting value of H$_{0}$ is the true cosmic value of
H$_{0}$, because the Hubble line holds out to $20,000\kms$ and can be
extended to $z>1$ with overlapping data from several SN\,Ia
collections \citep[e.g.,][]{Hicken:etal:09,Kessler:etal:09}.  

     The SN \textsl{HST} Project provided Cepheids in eight
SNe\,Ia-hosting galaxies. Revised period-color and P-L relations
allowing for metallicity differences (provided in their latest 
form for fundamental and overtone pulsators in \citealt{TRS:11}) were
applied to derive Cepheid distances \citep{Saha:etal:06}. 
Combining these distances in the summary paper of the SN \textsl{HST}
Project, including two additional galaxies from external sources, with
the apparent $m_{V}^{\rm corr}$ magnitudes, yields a value of 
$\langle M_{V}^{\rm corr}\rangle=-19.46\pm0.07\mag$ and hence with
Eq.~(\ref{eq:01}), H$_{0}=62.3\pm1.3\pm5.0\ksm$ \citep{STS:06}.

     The \textsl{HST} Key Project found, using the Cepheid distances
of eight SNe\,Ia and barely in statistical agreement, 
H$_{0}=71\pm2\pm6\ksm$ \citep{Freedman:etal:01}, whereas
\citet{Riess:etal:11} found from six SNe\,Ia a significantly higher
value of H$_{0}=73.8\pm2.4\ksm$. The divergence of the results depends
almost entirely on the different treatment of Cepheids. The latter
authors assume the P-L relation of the LMC to be universal and that
all color differences of Cepheids with equal periods are caused by
internal absorption.  

     Cepheids are indeed complex distance indicators. Their colors and
luminosities depend on metallicities, which are under revision
\citep{Bresolin:11,Kudritzky:Urbanaja:12}, and on the disentangling of
metallicity and internal-absorption effects. There are unexplained
differences within a given galaxy, and some excessively blue Cepheids
suggest the effect of an additional parameter, possibly the Helium
content \citep{TR:12a}. Plans to use infrared magnitudes
of Cepheids for the SN\,Ia calibration may alleviate some of the
problems, but it is obvious that the Cepheid-based calibration 
needs independent confirmation.

\section{The Tip of the Red-Giant Branch (TRGB) as a Distance Indicator}
\label{sec:6}
Sandage turned to a new distance indicator during his last years: the
tip of the red-giant branch (TRGB) as observed in galaxian halos. A
brief history is given elsewhere \citep{TR:12b}. 
The physical background is that old, metal-poor stars 
terminate---independent of mass---their evolution up the red-giant
branch by a helium flash in their electron-degenerate cores
(see \citealt{Salaris:12} and references therein).  

     The TRGB $I$-band magnitude, $M_{I}^{*}$, has exceptionally
favorable properties as a distance indicator: its physics is well
understood, being observed in outer halo fields it suffers virtually
no internal absorption and blends, and as a cut-off magnitude it is
free of selection effects; moreover it varies little over a wide
metallicity range ($-2.0<$[Fe/H]$<-1.2\dex$). The calibration of the
TRGB magnitude $M_{I}^{*}$ is very solid. \citet{Sakai:etal:04} found
$M_{I}^{*}=-4.05\mag$ from globular clusters. \citet{Rizzi:etal:07}
fitted the horizontal branch (HB) of five galaxies to a
metal-corrected HB from \textsl{Hipparcos} parallaxes and obtained the
same value. The 24 galaxies with known RR\,Lyrae distances and known
TRGB magnitudes yield $M_{I}^{*}=-4.05\pm0.02\mag$ with a dispersion
of only $0.08\mag$; the underlying RR\,Lyrae luminosity of 
$M_{V}({\rm RR})=0.52\mag$ at [Fe/H]$=1.5\dex$ \citep{ST:06}
is now robustly confirmed by \citet{Federici:etal:12}. The models of
\citet{Salaris:12} also give the same value of $M_{I}^{*}$. A value of
$M_{I}^{*}=-4.05\pm0.05\mag$ is adopted in the following.   

     A practical problem of the TRGB as a distance indicator are star
fields that do not only contain a predominantly old halo population,
but that include also an important fraction of Population~I stars. In
the latter case, evolved asymptotic giant-branch (AGB) stars and
supergiants, which may become as red and even brighter than the TRGB,
may swamp the RGB and make the detection of the true TRGB difficult or
impossible. Spurious detections are the consequence. The cases of
NGC\,3368, NGC\,3627, and NGC\,4038 are discussed in \citet{TR:12b}.
While this paper was written, the ambiguity of M101 has been solved by
\citet{Lee:Jang:12}; they determined the TRGB in eight galaxy fields,
yielding a high-precision mean apparent magnitude of
$m^{*}=25.28\pm0.01\mag$.

\section{The Local Velocity Field}
\label{sec:7}
\begin{figure}
\begin{center}
\begin{minipage}[t]{.58\linewidth}
   \vspace{0pt}
   \hspace*{-0.6cm}
   \includegraphics[width=1.00\textwidth]{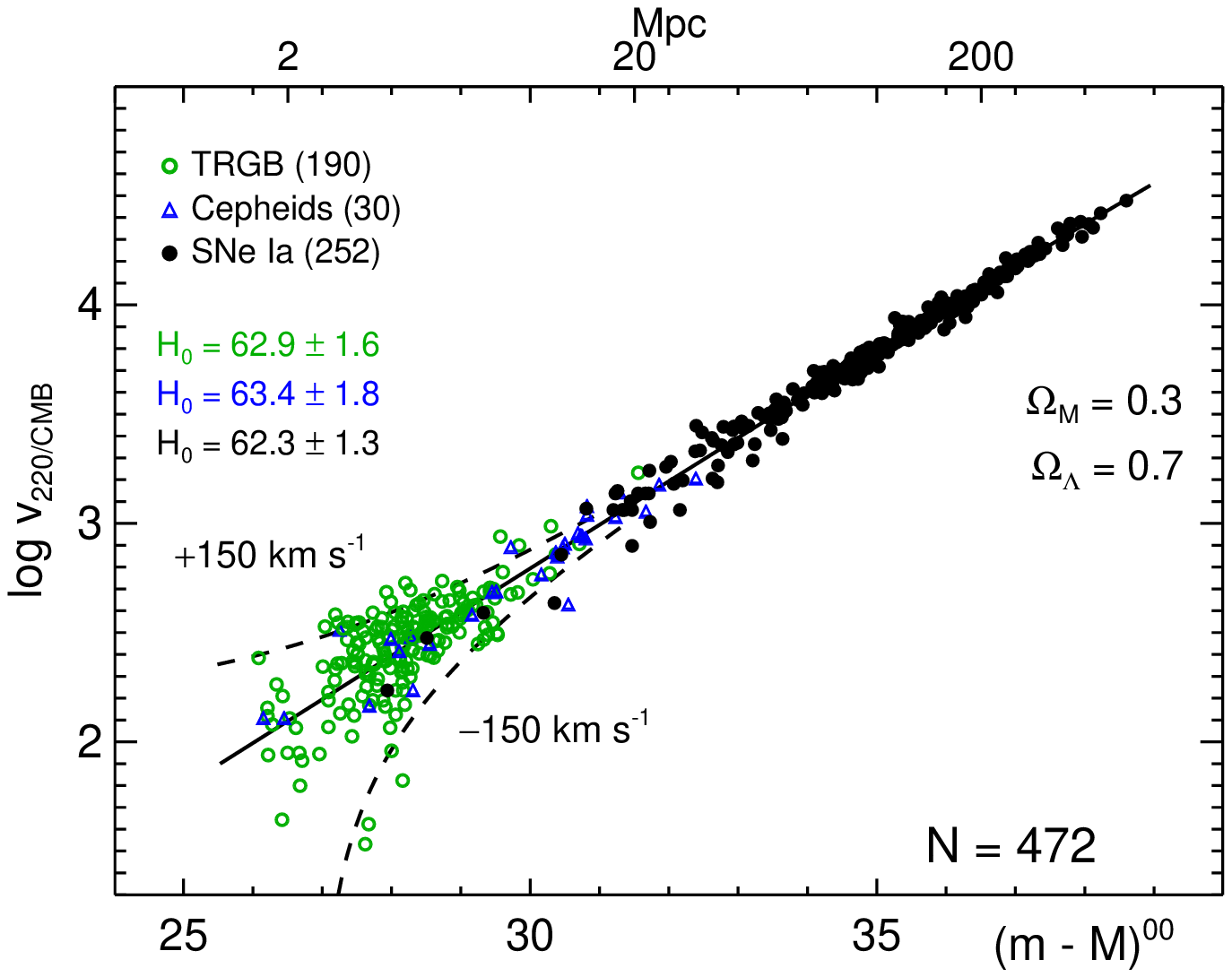}
\end{minipage}
\begin{minipage}[t]{.31\linewidth}
   \vspace{18pt}
   \hspace*{12pt}
   \includegraphics[width=1.00\textwidth]{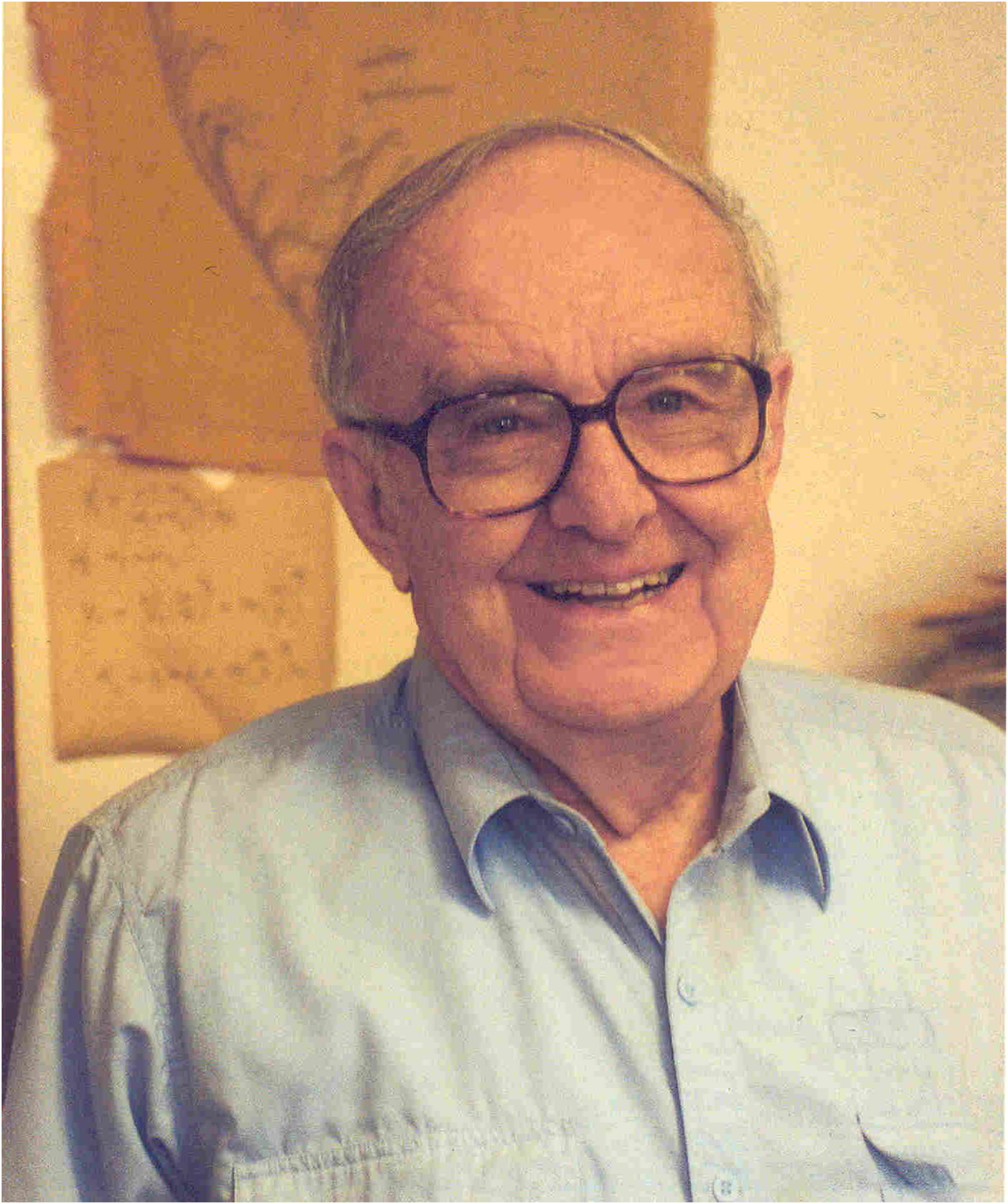}
\end{minipage}
\end{center}
   \caption{%
     (left)
     Distance-calibrated Hubble diagram showing distance moduli
     from the TRGB (green open circles), Cepheids (blue triangles),
     and SNe\,Ia (black dots). The slightly curved Hubble line holds
     for a model with $\Omega_{M}=0.3$, $\Omega_{\Lambda}=0.7$. The
     envelopes for peculiar velocities of $\pm150\kms$ are shown as
     dashed lines.  
     (right)
     Allan Sandage in front of his CMD of open clusters, ca. 1990. 
   }   
   \label{fig:07}
\end{figure}

TRGB magnitudes $m_{I}^{*}$ of over 200 galaxies are available in the
literature, 190 of them lie outside the Local Group (for a compilation
see, e.g., \citeauthor{TSR:08b} [\citeyear{TSR:08b}], with some
corrections and additions by various authors). Their distances are
derived from the above calibration, $M_{I}^{*}=-4.05\mag$, and are 
expressed as distance moduli $(m-M)^{00}$ from the barycenter of the
Local Group, assumed to lie two thirds of the way toward M31. The 190
galaxies are plotted in a Hubble diagram in Fig.~\ref{fig:07}. 
The velocities $v_{220}$ are corrected for a  self-consistent
Virgocentric infall model with a local infall vector of $220\kms$ and
a density profile of the Local Supercluster of $\rho\sim r^{-2}$
\citep{Yahil:etal:80}; the correction $\Delta v_{\rm Virgo}$ follows
then from eq.~(5) in \citet{STS:06}.   

     The scatter in Fig.~\ref{fig:07} increases with decreasing
distance as a result of the peculiar velocities of individual
galaxies, of order $50-70\kms$. An excess of slow galaxies at very
short distances is obviously the result of the pull from the Local
Group. The 79 TRGB galaxies with $(m-M)>28.2\mag$ define a Hubble line
with slope $0.199\pm0.019$, in agreement with a locally constant
expansion rate. The mean value of H$_{0}$ is well-defined at
$62.9\pm1.6\ksm$ (statistical error) at a median velocity of
$v_{220}=350\kms$.

     Also plotted in Fig.~\ref{fig:07} are 34 galaxies (outside the
Local Group) with Cepheid distances from \citet{Saha:etal:06}. The 30
galaxies with $(m-M)>28.2\mag$ yield the expected Hubble line slope of
$0.200\pm0.010$ and a mean value of H$_{0}=63.4\pm1.8\ksm$ at a median
velocity of $v_{220}=900\kms$.  

     In addition, Fig.~\ref{fig:07} shows 252 SNe\,Ia with known
maximum magnitudes $m_{V}^{\rm corr}$ in the system of \citet{RTS:05}
and with $v<30,000\kms$. Their moduli follow from the calibration
$M_{V}^{\rm corr}=-19.46\mag$ in Section~\ref{sec:5}. The 190 SNe\,Ia
within $3000<v<20,000\kms$ fit the Hubble line for $\Omega_{M}=0.3$
and $\Omega_{\Lambda}=0.7$ with a scatter of only $0.15\mag$. The
SNe\,Ia give a large-scale value of H$_{0}=62.3\ksm$, i.e.\ the same
as the more restricted SN sample in Section~\ref{sec:5}.   

     The agreement within statistics of H$_{0}$ from Cepheids and
SNe\,Ia is by construction, because the SN luminosities are calibrated
by means of a subsample of these Cepheids. However, the close
agreement of H$_{0}$ from Cepheids and SNe\,Ia with the mean value of
H$_{0}$ from local and \textit{independent} TRGB distances is highly
significant, indicating that any change of H$_{0}$ is undetectable
over a range as wide as  $300<v<30,000\kms$. An equivalent conclusion
can be drawn also without any absolute distance scale, because TRGBs,
Cepheids, and SNe\,Ia have sufficient redshift overlap to smoothly
connect the three segments of the Hubble line into a single line. This
limits the change of H$_{0}$ to 4\% over the entire distance range
\citep{TR:12a}. 

\begin{figure}
\begin{center}
\begin{minipage}[t]{.40\linewidth}
   \vspace{0pt}
   \hspace*{-0.3cm}
   \includegraphics[width=1.00\textwidth]{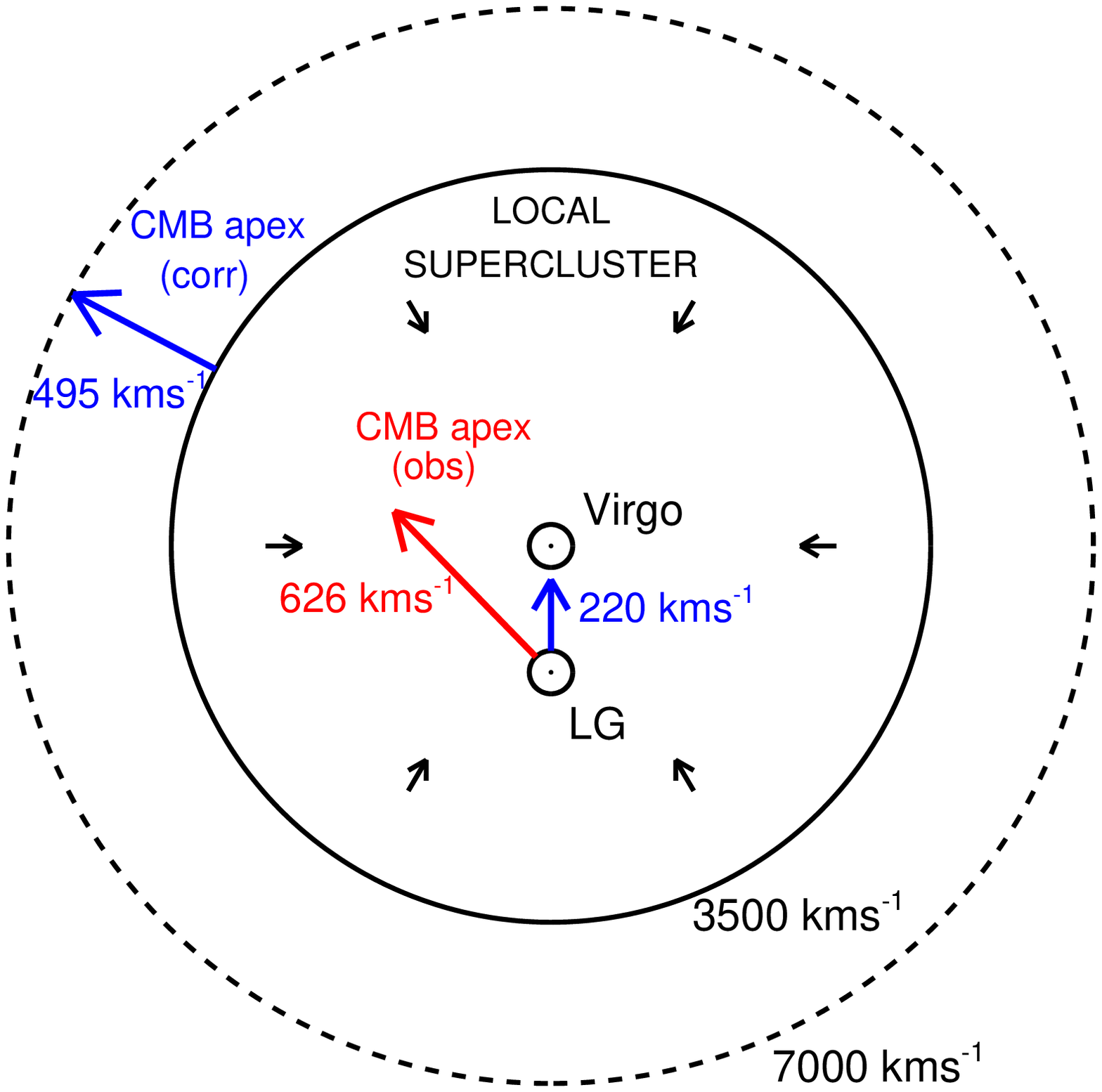}
\end{minipage}
\begin{minipage}[t]{.45\linewidth}
   \vspace{20pt}
   \hspace{0.9cm}
   \includegraphics[width=1.00\textwidth]{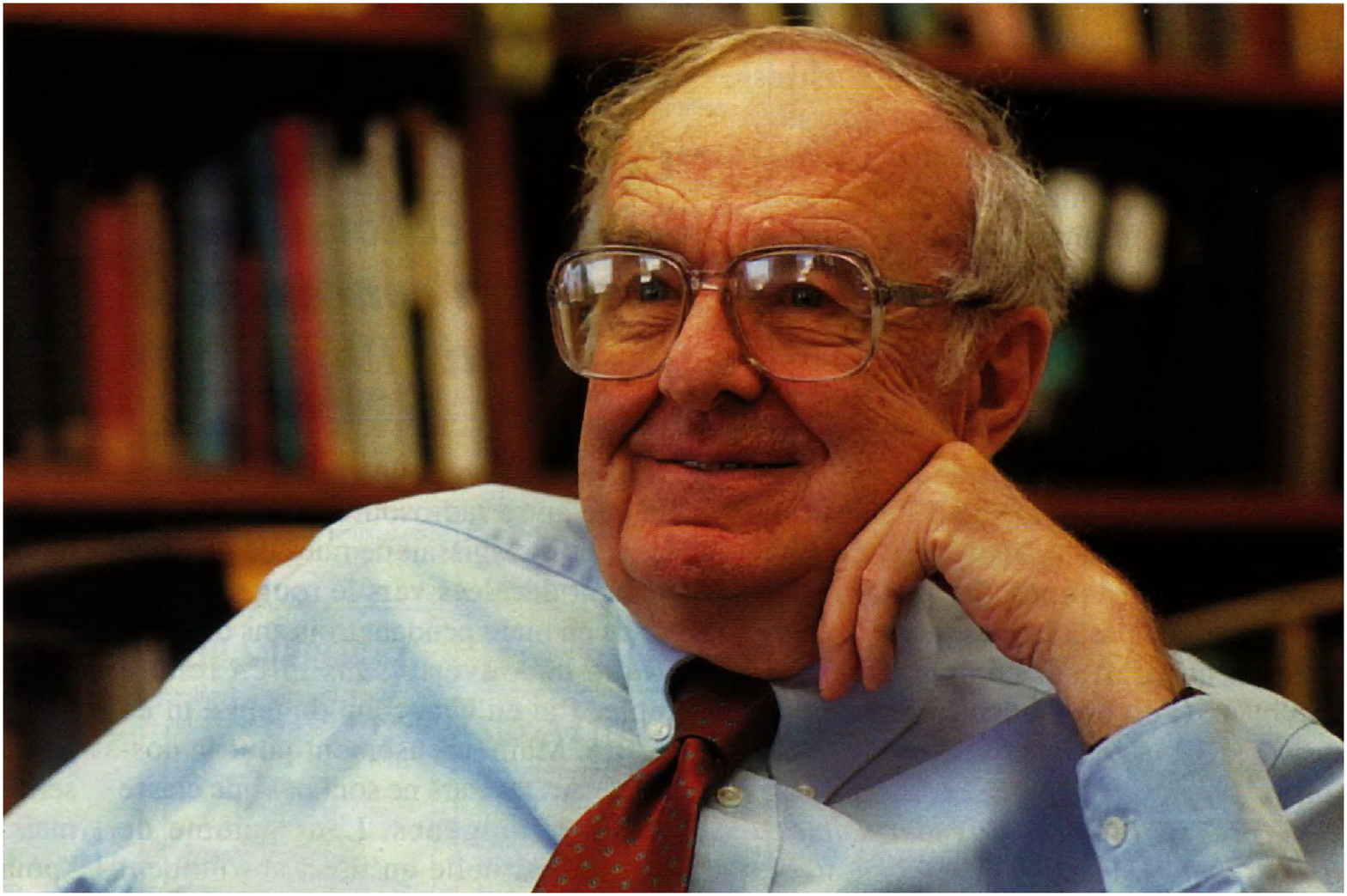}
\end{minipage}
\end{center}
   \caption{%
     (left)
     A schematic view of the Local Supercluster with the Virgo
     cluster in its center. The off-center Local Group (LG) is shown
     with its Virgocentric infall vector. The observed local velocity
     vector relative to the CMB is shown as well as the bulk motion,
     corrected for the Virgocentric infall vector, of the Local
     Supercluster toward the corrected apex $A_{\rm corr}$.  
     (right)
     Allan Sandage in 1998. \textit{(Photo: Ciel et Espace)}
}   
   \label{fig:08}
\end{figure}

\begin{figure}
\begin{center}
\begin{minipage}[t]{.47\linewidth}
   \vspace{0pt}
   \hspace*{-0.5cm}
   \includegraphics[width=1.00\textwidth]{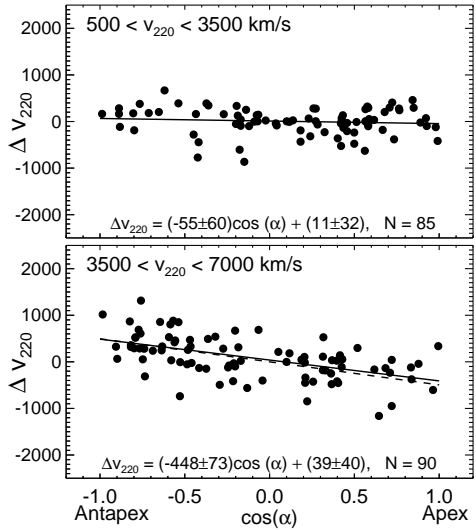}
\end{minipage}
\begin{minipage}[t]{.40\linewidth}
   \vspace{22pt}
   \centering
   \hspace*{21pt}
   \includegraphics[width=1.00\textwidth]{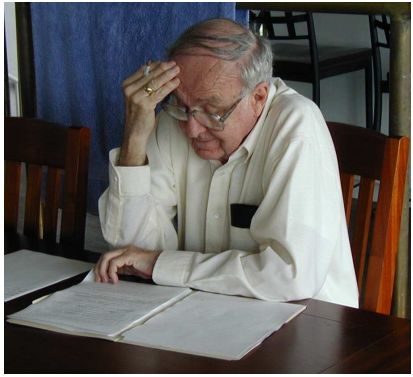}
\end{minipage}
\end{center}
   \caption{%
     (left)
     The velocity residuals $\Delta v_{220}$ from the Hubble line in
     Fig.~\ref{fig:07} shown in function of $\cos(\alpha)$, where
     $\alpha$ is the angle between the object and the corrected CMB
     apex $A_{\rm corr}$. a) for objects with $500<v_{220}<3500\kms$,
     and b) for objects with $3500<v_{220}<7000\kms$.  
     (right)
     Allan Sandage in 2002.
}
   \label{fig:09}
\end{figure}

     The data sets in Fig.~\ref{fig:07} are well suited to shed some
light on the motion of the local Volume causing the cosmic microwave
background (CMB) dipole. Once the observed apex velocity and direction
\citep{Hinshaw:etal:07} are corrected for the local Virgocentric
infall vector of $220\kms$, one  obtains a predicted non-Hubble
velocity of $495\pm25\kms$ in the direction of a corrected apex
$A_{\rm corr}$ at $l=275\pm2$, $b=12\pm4$ degrees in the constellation
of Vela \citep[Fig.~\ref{fig:08}; see also][]{ST:84}. The question is
how large is the co-moving volume of the Local Supercluster? 
An answer is given in Fig.~\ref{fig:09}, where the residuals 
$\Delta v_{220}$ from the Hubble line in Fig.~\ref{fig:07} are plotted
against $\cos(\alpha)$; $\alpha$ is the angle between a given
object and $A_{\rm corr}$.  
The flat distribution of the objects with $500<v_{220}<3500\kms$ in
Fig.~\ref{fig:09}a rules out any systematic motion toward
$A_{\rm corr}$ within the Local Supercluster. In sharp contrast,
objects with $3500<v_{220}<7000\kms$ reveal in Fig.~\ref{fig:09}b a
highly significant (three-dimensional) velocity of $448\pm73\kms$ in
the direction of $A_{\rm corr}$ and in statistical agreement with the
predicted value of $495\pm25\kms$. The emerging picture is that the
Local Supercluster is a contracting entity, as strongly suggested by
the local Virgocentric infall, which moves in bulk motion relative to
the objects in a shell between $3500$ and $7000\kms$, constituting the
Machian frame. The acceleration of the Local Supercluster must be
caused by the irregular mass and void distribution within this shell.
The role of the Great Attractor as accelerator is not clear; it lies
with $v\sim4700\kms$ in the expected distance range, but $40^{\circ}$
away from $A_{\rm corr}$. In any case, Shapley's Supercluster at
$50^{\circ}$ from $A_{\rm corr}$ and at $v\sim13,000\kms$ is too
distant to contribute noticeably to the acceleration of the Local
Supercluster.  

     All galaxies with $v_{220}>3500\kms$ have been corrected in this
paper by $\Delta v_{\rm CMB}=495\cos(\alpha)$ to compensate for the
motion of the Local Supercluster relative to the CMB. 

\section{The Luminosity Calibration of SNe\,Ia from the TRGB}
\label{sec:8}
The first attempts to calibrate the SN\,Ia luminosity based upon TRGB
distances are from \citet{TSR:08a} and \citet{Mould:Sakai:09}. 
Then a year ago, and two years after Allan Sandage's death, the 
Stone of Rosetta appeared in the form of the unreddened
\textit{standard} SN\,Ia 2011fe \citep{Nugent:etal:11} in M101 with a
firm TRGB distance (see Section~\ref{sec:6}).
This brings the number of SNe\,Ia with known TRGB distances to six,
which forms a solid basis for a luminosity calibration of SNe\,Ia. The
SNe\,Ia are individually discussed elsewhere \citep{TR:12b}; their
relevant parameters (including the revised TRGB distance of M101 in
Section~\ref{sec:6}) and their compounded statistical errors are
compiled in Table~\ref{tab:01}. Their resulting weighted mean
luminosity of $M_{V}^{\rm corr}=-19.39\pm0.05\mag$, 
inserted in Eq.~(\ref{eq:01}), yields  
\begin{equation}
   {\rm H}_{0}=64.6\pm1.6\pm2.8\ksm,
\label{eq:02}
\end{equation}
where the systematic error is justified in \citet{TR:12b}.

     Cepheid distances are available for all six SNe\,Ia in
Table~\ref{tab:01}; they yield $M_{V}^{\rm corr}=-19.40\pm0.06\mag$
\citep[][their table~3]{TR:12b}, in fortuitous agreement with the TRGB
calibration. Additional weight to the present calibration is given by
two galaxies in the Fornax Cluster that have produced four SNe\,Ia
(SNe\,1980N, 1981D, 1992A, and 2006dd) and whose surface brightness
fluctuation (SBF) distances from \textsl{HST}/ACS are given by
\citet{Blakeslee:etal:10}; they lead, again independently, to a
weighted mean value of $M_{V}^{\rm corr}=-19.43\pm0.06\mag$.  

\begin{table}
\begin{center}
\caption{The TRGB calibration of SNe\,Ia}
\label{tab:01}  
\footnotesize
\begin{tabular}{llrccc}
\noalign{\smallskip}
\noalign{\smallskip}
\hline
\hline
\noalign{\smallskip}
\noalign{\smallskip}
   \multicolumn{1}{c}{SN}                          &
   \multicolumn{1}{c}{Gal.}                        &
   \multicolumn{1}{c}{$m^{\rm corr}_{V}$}          &
   \multicolumn{1}{c}{$(m\!-\!M)_{\rm TRGB}$}      &
   \multicolumn{1}{c}{Ref.}                        &
   \multicolumn{1}{c}{$M^{\rm corr}_{V}$}          \\
   \multicolumn{1}{c}{(1)}                         &
   \multicolumn{1}{c}{(2)}                         &
   \multicolumn{1}{c}{(3)}                         &
   \multicolumn{1}{c}{(4)}                         &
   \multicolumn{1}{c}{(5)}                         &
   \multicolumn{1}{c}{(6)}                         \\[-4pt]
\noalign{\smallskip}
\hline
\noalign{\smallskip}
 2011fe & N5457   & $ 9.93\,(06)$ & $29.33\,(02)$ & 1   & $-19.40\,(06)$ \\
 2007sr & N4038   & $12.26\,(13)$ & $31.51\,(12)$ & 2   & $-19.25\,(18)$ \\
 1998bu & N3368   & $11.01\,(12)$ & $30.39\,(10)$ & 3   & $-19.38\,(16)$ \\
 1989B  & N3627   & $10.94\,(11)$ & $30.39\,(10)$ & 3   & $-19.45\,(15)$ \\
 1972E  & N5253   & $ 8.37\,(11)$ & $27.79\,(10)$ & 4,5 & $-19.42\,(15)$ \\
 1937C  & I4182   & $ 8.92\,(16)$ & $28.21\,(05)$ & 4,5 & $-19.29\,(17)$ \\
\hline
\multicolumn{5}{l}{straight mean}             & $\quad-19.37\pm0.03$ \\
\multicolumn{5}{l}{weighted mean}             & $\quad-19.39\pm0.05$ \\ 
\hline
\end{tabular}
\end{center}
   (1) \citealt{Lee:Jang:12};
   (2) \citealt{Schweizer:etal:08};
   (3) galaxy assumed at the mean TRGB distance of the Leo\,I group;
   (4) \citealt{Sakai:etal:04};
   (5) \citealt{Rizzi:etal:07}.
\end{table}

\section{Results and Conclusions}
\label{sec:9}
The large-scale value of H$_{0}$ is provided by the Hubble diagram of
SNe\,Ia with $3000<v<20,000\kms$ (and beyond) pending the calibration
of their absolute magnitude and its error. Two nearby calibrations
based on young Population~I (Cepheids) and old Population~II (TRGB)
distance indicators lead, in good agreement, to a weighted mean SN\,Ia
luminosity of $M_{V}^{\rm corr}=-19.41\pm0.04\mag$. The result is
further and independently supported by four SNe\,Ia in two Fornax
galaxies with modern SBF distances. A secular change of the SN\,Ia
luminosity is very unlikely because it would have to change sign at
$z\sim0.9$. A combination of the adopted calibration with
Eq.~(\ref{eq:01}) leads therefore to a firm cosmic value of the Hubble
constant of H$_{0}=64.1\pm2.4\ksm$ (systematic error included).

     The calibrated SNe\,Ia are well suited to assess the distances to
many galaxies \citep[e.g.,][]{TSR:08b}, and particularly to clusters
with multiple occurrences. Prime example is the Fornax Cluster, with
five SNe\,Ia (as above as well as SN\,2001el) that give---with 
$\langle m_{V}^{\rm corr}\rangle=12.19\pm0.09\mag$---a distance of 
$(m-M)=(m-M)^{00}=31.60\pm0.10\mag$ (H$_{\rm Fornax}=66\pm4\ksm$), in
good agreement with the entirely independent mean SBF cluster distance
modulus of $31.55\pm0.04\mag$ \citep{Blakeslee:etal:10}.
The Virgo Cluster would need more than its present four SNe\,Ia for a
good distance determination because of its important depth effect. The
best value of $(m-M)=31.18\pm0.10\mag$ ($(m-M)^{00}=31.22\pm0.10\mag$;
H$_{\rm Virgo}=66\pm5\ksm$) comes from the difference between Fornax
and Virgo of $\Delta(m-M)=0.42\pm0.02\mag$, based on a wealth of SBF
distances \citep{Blakeslee:12}.

     A promising development is the extension of water megamaser
distances out to \linebreak[4]$\sim\!10,000\kms$. So far, four sources
give H$_{0}=70.5\pm5.4\ksm$ (Braatz, this volume), which is
statistically not excluded by the value derived here. Yet, at least in
one case, the value of H$_{0}$ depends heavily on the H$_{0}$ prior
chosen \citep{Reid:etal:12}.

     For values of H$_{0}$ from strong gravitational lenses and from
the Sunyaev-Zel'dovich effect, both quite model-dependent, the reader
is referred to Suyu (this volume) and Bonamente (this volume),
respectively.

     The CMB fluctuation spectrum (\textit{imprinted at an early epoch}) 
has led to many estimates of the (\textit{present}) value of H$_{0}$,
but they are necessarily model-dependent and rely on a variety of free
parameters and, in some cases, on the choice of priors.
\citet{Komatsu:etal:11} derived, from a simple six-parameter
analysis of the \textsl{WMAP7} data, H$_{0}=70.3\ksm$, yet imposing a
prior H$_{0}=74.2\pm3.6\ksm$. Other authors have included additional
evidence from large red galaxies (LRG), the shape of the Hubble
diagram of SNe\,Ia, and from baryon acoustic oscillations (BAO); their 
results cluster around H$_{0}=69\ksm$ 
\citep{Anderson:etal:12,Reid:etal:12a,Sanchez:etal:12}.

     \citet{Calabrese:etal:12} analyzed the \textsl{WMAP7} data,
combined with ground-based observations, at arcminute angular scales,
making different model assumptions. If they impose the standard value
of $N_{\rm eff}=3.046$ for the number of effective relativistic
neutrinos and set a prior of H$_{0}=68\pm2.8\ksm$, they find
H$_{0}=66.8\pm1.8\ksm$. In case of non-standard decoupling of
neutrinos, the value could be lower (A.~Melchiorri, priv. commun.).  

     Of particular interest is the BAO peak found in the correlation
function of the 6dF Galaxy Survey at a redshift of $z=0.106$, the 
lowest redshift so far (\citealt{Beutler:etal:11}; Colless, this
volume). This allows the authors to evaluate H$_{0}$, making minimum
use of CMB data. Their result of H$_{0}=67\pm3.2\ksm$ does not decide
yet between the present result and H$_{0}\sim70\ksm$, but it makes
still higher values in the literature less probable. Tighter
constraints are foreseeable.  

     Sandage's last published value of the Hubble constant is 
H$_{0}=62.3\ksm$ \citep{TS:10}. The value has here been slightly
revised to H$_{0}=64.1\pm2.4 \ksm$ to account for the high-weight TRGB
calibration of SNe\,Ia. If confirmed, the value will be useful to
constrain some of the fundamental parameters of the Universe. \\

\noindent
\textit{Dedication}: To the memory of a great man, Allan Sandage.\\

\begin{minipage}[t]{.99\linewidth}
   \centering
   \includegraphics[width=0.56\textwidth]{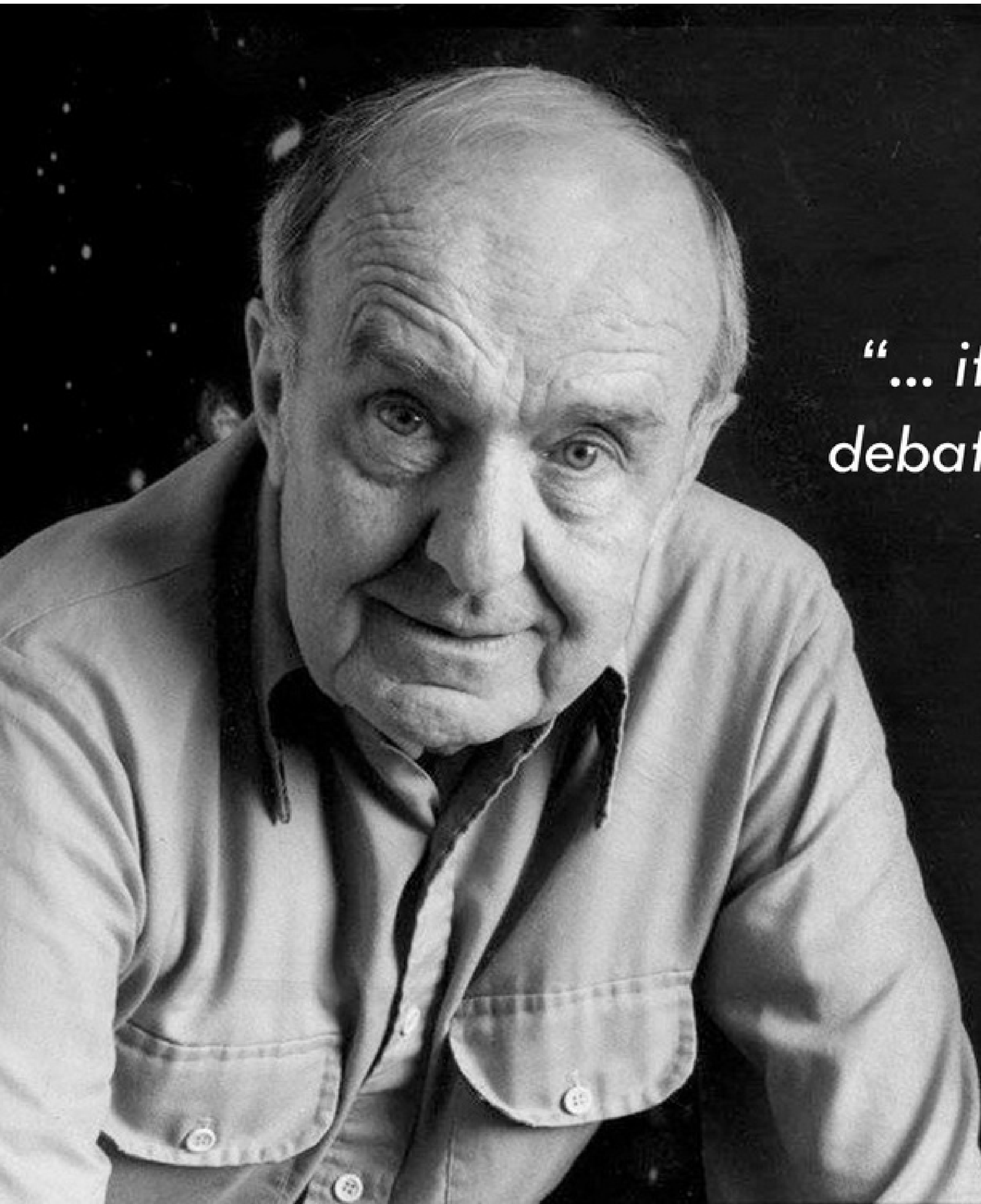}
\end{minipage}


\end{document}